\RequirePackage[colorlinks,allcolors=blue]{hyperref}
\documentclass{agujournal2019}
\usepackage{url}
\usepackage{soul}
\usepackage{amsmath}
\usepackage{amssymb}

\newcommand{\vq}{\mathbf q}
\newcommand{\vu}{\mathbf u}
\newcommand{\vx}{\mathbf x}
\newcommand{\va}{\mathbf a}

\begin{document}

\title{Dispersion and mixing in heterogeneous compressible porous media under transient forcing}

\authors{Satoshi Tajima\affil{1}, Marco Dentz\affil{2}}

\affiliation{1}{Graduate School of Frontier Sciences, The University of Tokyo, Kashiwa, Japan}
\affiliation{2}{Spanish National Research Council (IDAEA-CSIC), Barcelona, Spain}

\correspondingauthor{Satoshi Tajima}{tajima@envsys.k.u-tokyo.ac.jp}

\begin{abstract}
Periodic forcing of flow in compressible porous media is an important driver for
solute dispersion and mixing in geological and engineered porous media subject
for example to tides, pumping and recharge cycles, or fluid injection and
withdrawal cycles with a wide range of environmental and industrial
applications. The combination of periodic forcing, spatial medium heterogeneity
and medium compressibility leads to intricate spatio-temporal flow, dispersion
and mixing patterns. We analyze these patterns using detailed numerical
simulations based on a stochastic representation of the spatial medium
heterogeneity. Solute dispersion is characterized by the interface length and
width, mixing in terms of the dilution index and the distribution of
concentration point values. Poincar\'e maps show how the interplay of
heterogeneity and compressibility leads to the creation of stable regions that
inhibit the advancement and dispersion of the mixing interface, and chaotic
regions that at the same time enhance solute mixing. This means that spatial heterogeneity in combination with temporal forcing can lead to the containment
of solute and at the same time promote mixing.
\end{abstract}

\section{Introduction}
Transient forcing of flow drives solute mixing and dispersion in compressible media through the interplay of temporal fluctuations, medium compressibility,
and spatial heterogeneity. It plays a key role in mixing processes in
geological and engineered porous media in response to tidal fluctuations, due to
fluid injection and withdrawal cycles, and periodic pumping and recharge with
broad environmental and engineering applications. For example, the salinity
distribution in coastal aquifers controlled by transient forcing including tidal
fluctuations~\cite{Oberdorfer1990, Inouchi1990, Pool2014, Ataie-Ashtiani1999}
and infrequent events such as storm surges~\cite{Terry2010, Bailey2014,
  Tajima2023}. In the context of underground hydrogen storage, hydrogen purity
is determined by the mixing with the reservoir fluid~\cite{hydrogenreview}, which is enhanced
by injection-production cycles.

Porous media have two primary characteristics that influence transient transport
dynamics. First, porous media inherently exhibit heterogeneity across spatial
scales. At the Darcy scale, such a heterogeneous structure is represented by
heterogeneous hydraulic conductivity fields~\cite{Dagan2012, Dentz2023}. Second,
porous media are often compressible, which is quantified by a finite specific
storage. This parameter denotes the volume of water released per unit decline in
the hydraulic head, accompanied by a decrease in pore water pressure and an
increase in effective stress. Finite storage attenuates the amplitude of
periodic flow propagating into porous media and delays its
phase~\cite{Jacob1950, Jiao1999, Li2000, Li2002, Nielsen1990}.

For incompressible porous media, perturbation analyses and numerical simulations
have uncovered that velocity fluctuations in the mean flow direction minimally
affect longitudinal dispersion parallel to the mean flow within heterogeneous
porous media, whilst exerting notable influence on transverse dispersion in the
orthogonal direction~\cite{kinzelbach1986modelisation, Rehfeldt1992, Dagan1996, Dentz2003,
  Cirpka2003, deDreuzy2012}. In highly heterogeneous porous media, temporal
fluctuations in velocity reduce longitudinal spreading compared with
steady-state conditions~\cite{deDreuzy2012}.

\citeA{Pool2016} study the effects of finite storage on the evolution of
displacement fronts under transient forcing in homogenous porous media using
laboratory experiments, numerical simulations, and derive explicit analytical
expression for the time-average transport dynamics.
The combined effects of spatial heterogeneity and finite storage on dispersion
were investigated by~\citeA{Pool2015}, focusing on
seawater intrusion under tidal fluctuations. Their numerical simulations
reveal that increasing heterogeneity mitigates the spreading of the solute
interface induced by periodic flow, while increased storage augments spreading.
\citeA{pool2018effects} study the combined effect of medium compressibility,
heterogeneity and variable density on reactive mixing patterns under temporal
flow fluctuations. Their numerical simulations provide evidence that interface deformation due to spatial heterogeneity and temporal flow fluctuations together with density variation leads to the formation of complex patterns of reaction
hotspots. It is still not clear, however, how medium heterogeneity and
compressibility together with temporal flow
fluctuations control the occurrence of Lagrangian coherent structures, and how they affect the advancement, dispersion and mixing dynamics of a solute
interface aligned with the fluctuating pressure boundary. 
Recently, \citeA{lester2024} shed light on how Lagrangian coherent structures
control solute dispersion in heterogeneous poroelastic media. To this end,
these authors analyze the dispersion of point-like initial plumes under temporal fluctuations in the presence of a mean
flow toward the fluctuating pressure boundary. They use a diffusive particle
mapping method as a surrogate for the solution of the full advective-diffusive
transport problem.

In this paper, we focus on the dispersion and mixing of a solute front that invades the medium from the fluctuating pressure
boundary motivated by saltwater fronts in tidal aquifers, groundwater remediation
measures through the period injection of reactants, and geostorage operations such as underground hydrogen storage.  
To this end, we first consider purely advective transport to elucidate
the interplay between flow
patterns and fluid deformation. Poincar\'e maps of the kinematic equations of
motion of fluid particles in the flow field highlight the creation of stable
(stagnant) and chaotic regions and their impact on large-scale solute
dispersion. Detailed random walk particle tracking simulations quantify the
impact of local-scale dispersion on the evolution of mixing fronts in terms of
effective and ensemble dispersion coefficients, the mixing volume as measured by
the dilution index, and the evolution of the probability density function of
concentration point values. Our analysis sheds light on the intricate relation
between flow and mixing patterns under transient forcing.

\section{Basics and Methods}
\subsection{Flow}
\begin{figure}[htbp]
  \includegraphics[width=\linewidth, pagebox=artbox]{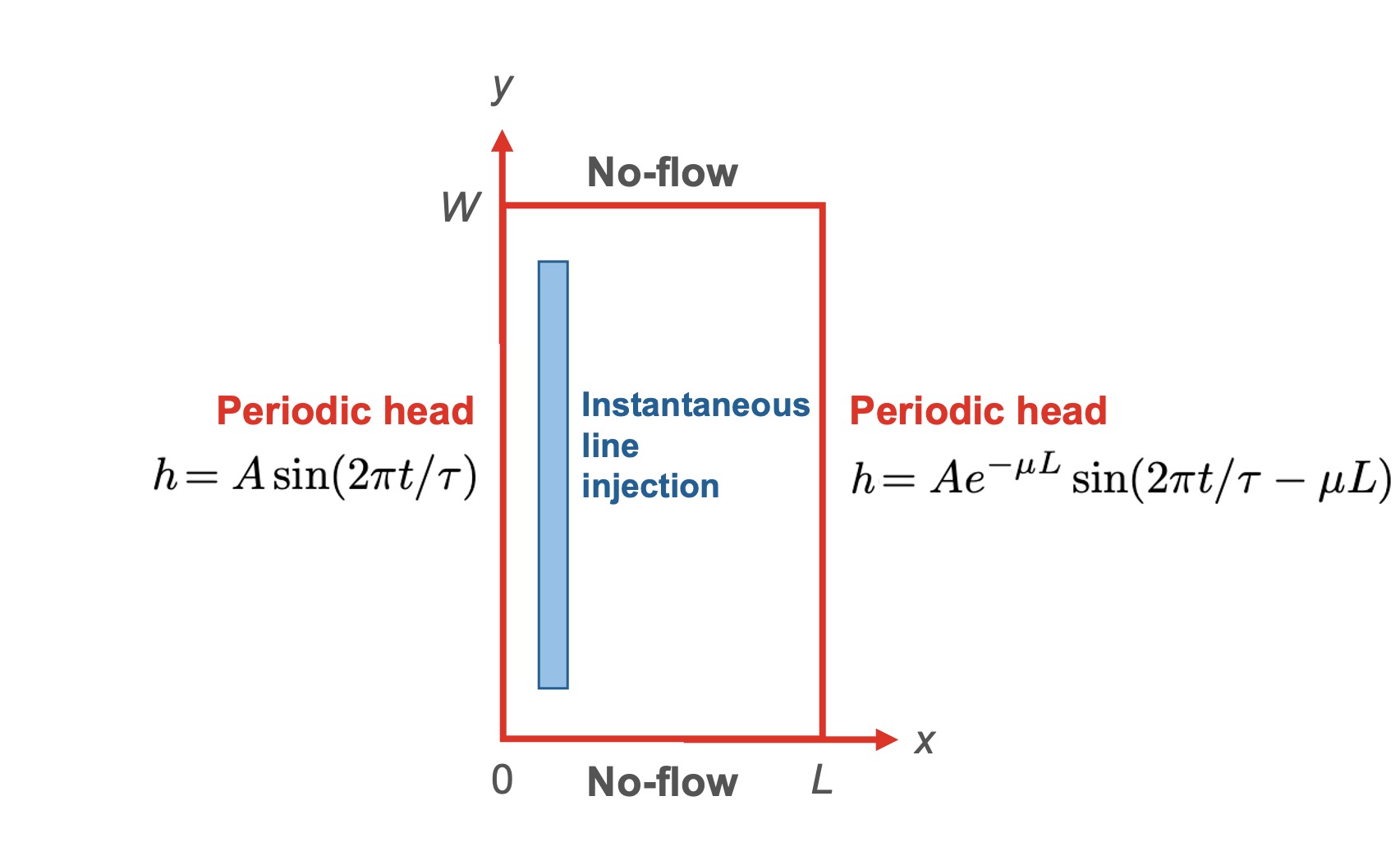}
  \caption{Schematic of model domain and boundary conditions.}
  \label{fig:setting}
\end{figure}

We consider transient flow and solute transport within a two-dimensional, fully
saturated, confined, heterogeneous porous medium (Figure~\ref{fig:setting}). Fluid mass and momentum
balance are expressed by~\cite{Bear1972} 

\begin{align}
S_s \frac{\partial h(\vx, t)}{\partial t} + \nabla \cdot \mathbf q(\vx,t) = 0,
&& \mathbf q(\vx,t) = - K(\vx) \nabla h(\vx, t),
  \label{eq:flow}
\end{align}

where $S_s$ is the specific storage, $t$ is the time, $h(\vx,t)$ is the
hydraulic head, $\vq(\vx, t)$ is the Darcy flux, and $K(\vx)$ is the
heterogeneous hydraulic conductivity. The heterogeneity of $K(\vx)$, assumed
scalar for simplicity, is modelled as a multi-Gaussian spatial random field. The
log-hydraulic conductivity $f(\vx) = \ln K(\vx)$ is represented as a correlated
Gaussian random space function with a mean $\overline{f} =$ constant and the exponential covariance function

\begin{equation}
\mathcal C_f(\vx) = \sigma_{f}^2 \exp \left(-{|\vx|}/{\lambda}\right)
  \label{eq:Kcov}
\end{equation}

where $\sigma_{f}^2$ is the variance of $f(\vx)$, $\lambda$ its correlation
length. The effective hydraulic conductivity is equal to the geometric mean
conductivity $K_G = \exp(\overline f)$ \cite{Renard:1997}.

We consider sinuisoidal head fluctuations at $x = 0$ with amplitude $A$ and
period $\tau$,

\begin{align}
    \left.h(\vx,t)\right|_{x = 0} =A \sin(2\pi t/\tau) \label{eq:head_BC}. 
\end{align}

For a semi-infinite equivalent homogeneous medium with $K(\vx) = K_G$, the analytical
solution of Eq~\eqref{eq:flow} is~\cite{Jacob1950,ferris1952cyclic}

\begin{align}
h_e(x,t) = A \exp(-\mu x) \sin(2\pi t/\tau-\mu x). 
\end{align}

The wave number $\mu$ is defined as

\begin{equation}
  \mu \equiv \sqrt{\frac{S_s \pi}{K_G \tau}}.
  \label{eq:mu}
\end{equation}

Its inverse gives the characteristic penetration distance of the periodic head
fluctuation, or in other words, the head fluctuation decays exponentially fast
on the length scale $1/\mu$. The corresponding Darcy flux is given by

\begin{equation}
  q_e(x,t) = q_0 \exp(-\mu x) \sin\left(\frac{2\pi t}{\tau} - \mu x + \frac{\pi}{4}\right),
 \label{eq:v_hom_ana}
\end{equation}
with the maximum flux

\begin{equation}
  q_0 = \sqrt{2} A K_G \mu.
  \label{eq:v0}
\end{equation}

\begin{figure}[htbp]
  \includegraphics[width=0.5\linewidth, pagebox=artbox]{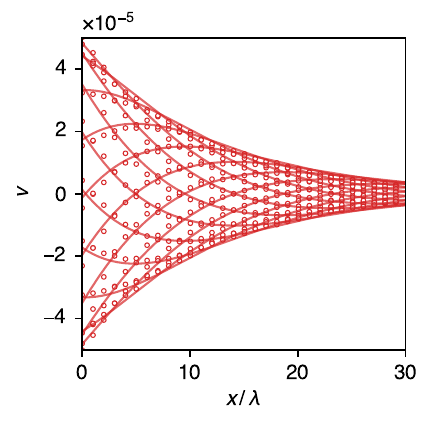}
  \caption{The symbols show the vertically averaged velocity $\overline v(x)$ at various times within one period for a heterogeneous medium with $\sigma_f^2 = 2.0$.
  The lines show the analytical solution for an equivalent homogeneous medium characterized by $K_G$.}
  \label{fig:het_v}
\end{figure}

Figure~\ref{fig:het_v} shows that the analytical solution~\eqref{eq:v_hom_ana} is an excellent descriptor of the vertically averaged streamwise flow velocity. 

Here we consider a finite two-dimensional medium of
width $W$ and length $L$.  At the horizontal domain boundaries, we specify
no-flow boundary conditions. At the right domain boundary at $x = L$, we set
$h(x = L,t) = h_e(x = L,t)$ to emulate flow in a semi-infinite medium. 
Equation~\eqref{eq:flow} is solved using MODFLOW
6~\cite{Langevin2017}. The simulation parameters are given in
Table~\ref{tab:parameters}. The domain length is at least $8/\mu$, that is, $8$
times the penetration depth.
Steady-state conditions are achieved after a simulation time of $t = 100 \tau$. 
Note that in the numerical transport simulations solute is injected when the
flow field is in a steady state, which sets the time $t = 0$ for transport.

\begin{table}[htbp]
  \caption{Parameters for numerical flow and transport simulations.}
  \label{tab:parameters}
  \centering
   \begin{tabular}{llll}
    \hline
    Parameter & Symbol & Value & Unit \\
    \hline
    Domain length & $L$ & $50 \lambda$ & m \\
    Domain width  & $W$ & $30 \lambda$ & m\\
    \hline
    Correlation length & $\lambda$ & 1 & m \\
    Geometric mean hydraulic conductivity & $K_G$ & $1\times10^{-4}$ & m s$^{-1}$ \\
    Log-conductivity variance & $\sigma_f^2$ & 0, 0.25, 0.5, 1.0, 1.5, 1.75, 2.0 & - \\
    Amplitude & $A$ & 1 & m \\
    Period & $\tau$ & 43200 & s \\
    Porosity & $\phi$ & 0.25 & - \\
    Specific storage & $S_s$ & $1\times10^{-2}$ & - \\
    \hline
    Initial line length & $\ell_0$ & 0.8 W & m\\
    Injection point & $x_0$ & 0.1 L & m \\
    Dispersivity & $\alpha$ & $1\times10^{-3}, 5\times10^{-3}$ & m \\
    P\'{e}clet number & $Pe$ & 200, 1000 & - \\
    \hline
   \end{tabular}
\end{table}

\subsection{Transport}

Transport is described by the advection-dispersion equation~\cite{Bear1972}

\begin{equation}
    \frac{\partial c(\vx, t)}{\partial t} = - \nabla \cdot \vu(\vx, t) c(\vx, t)
    + \nabla \cdot \left[\mathbf{D}(\vx,t) \nabla c(\vx,
    t)\right], \label{eq:ADE}
\end{equation}

where $c(\vx, t)$ is the mass concentration, $\vu(\vx, t) = \vq(\vx)/\phi$ is
the pore water velocity. Here, we assume that porosity variability with pressure
is small and set it equal to constant in the transport equation. Furthermore,
the density of water is assumed to be constant, that is, it does not depend on
the solute concentration. The local hydrodynamic dispersion tensor
$\mathbf{D}(\vx,t)$ is defined by~\cite{Bear1972}

\begin{align}
    D_{ij}(\vx,t) = D_0 \delta_{ij} + \delta_{ij} \alpha_I |\vu(\vx,t)| + (\alpha_I - \alpha_{II}) \frac{u_i(\vx,t)u_j(\vx,t)}{|\vu(\vx,t)|}. 
    \label{eq:D_def}
\end{align}

For simplicity, we assume that $D_0=0$ and $\alpha_I = \alpha_{II} \equiv
\alpha$. Thus, $D_{ij} = \delta_{ij} \alpha |\vu(\vx,t)|$. The relative strength
of advective and dispersive transport mechanisms is measured by the P\'eclet
number

\begin{equation}
    Pe \equiv \frac{\lambda}{\alpha}.
    \label{eq:Pe_def}
\end{equation}

For $Pe \gg 1$, advection dominates.
We consider an instantaneous injection along a line of length $\ell_0$ at $x = x_0$ at  time $t = 0$,

\begin{align}
    c(\vx, t=0) &= \frac{1}{\ell_0}\delta(x - x_0),
\end{align}

where $\delta(x)$ denotes the Dirac delta. At the left and right domain boundaries, we specify absorbing boundary conditions. Note that no mass exits the domain during the simulation time, and thus the solute concentration integrates into one at any time.  

The advection-dispersion equation~\eqref{eq:ADE} is equivalent to the Langevin
equation~\cite{KIB1988, Labolle2000},

\begin{equation}
    \frac{d \vx(t)}{dt} = \mathbf{u}[\vx(t)] + \nabla\cdot\mathbf{D}[\vx(t)] + \mathbf{B}[\vx(t)]\cdot\boldsymbol{\zeta}(t),
    \label{eq:Langevin}
\end{equation}

where $\mathbf{B}\cdot\mathbf{B} = 2 \mathbf{D}$, and $\boldsymbol{\zeta}(t)$ is a Gaussian white noise with zero mean and correlation function

\begin{equation}
    \langle\zeta_i(t)\zeta_j(t')\rangle = \delta_{ij} \delta(t - t'). 
\end{equation}

The angular brackets denote the average over all noise
realizations. We use here the Ito interpretation of the stochastic integral~\cite{Risken:1996}. 
The transport problem is solved numerically based on the Langevin
equation~\eqref{eq:Langevin} as outlined in detail in~\ref{app:numerical}. The concentration distribution $c(\vx,t)$ is given in terms of the particle positions $\vx(t)$ as

\begin{align}
c(\vx,t) = \left\langle \delta[\vx - \vx(t)] \right\rangle, 
\end{align}

where $\delta(\vx)$ denotes the Dirac delta distribution. To reconstruct the concentration values from the particle positions, we use kernel density estimators \cite{fernandez2011optimal} as detailed in~\ref{app:numerical}. 

Note that the velocity decreases exponentially with distance from the inlet on the length scale $\ell_c = 1/\mu$, the average penetration depth of the boundary pressure fluctuations. Thus, the interface starts decelerating when it reaches the distance $\ell_c$. The characteristic time for deceleration is given by \cite{Pool2016} 

\begin{align}
  \tau_v = \frac{2\pi\exp\left(2\mu x_{\mathrm i}\right)}{u_0^2 \mu^2 \tau}.
  \label{eq:tau_v_def}
\end{align}

If $\ell_c < \lambda$, the mixing front starts decelerating before it can sweep over a correlation length of the medium. In other words, it experiences the medium heterogeneity as a quasi-stratified medium. The scenarios under consideration here are characterized by $\ell_c > 5 \lambda$. Thus, the mixing interface can sweep more than one correlation length during a period $\tau$. The transport parameters used in the numerical simulations are detailed in Table~\ref{tab:parameters}. 

\subsection{Observables}
In the following, we define the observables used to characterize the dispersion and mixing behaviours. Dispersion is quantified in terms of the displacement statistics, specifically in terms of the displacement mean and variance, and the line length under purely advective transport.  Mixing is quantified in terms of the statistics of the concentration point values, which are used to determine the mixing volume in terms of the dilution index, and the probability density function of concentration point values. All observables $\omega(t)$ are time averaged over one period $\tau$ of the periodic pressure fluctuation,

\begin{equation}
   \overline{\omega}(t) \equiv \frac{1}{\tau}\int^{\tau}_0 dt' \omega(t+t').
\end{equation}
 
The time average is denoted by an overbar. The observables defined in the following are compared to the corresponding analytical solutions for equivalent homogeneous media given in \cite{Pool2016}. For the convenience of the reader, these solutions are summarized in~\ref{app:hom}. 

\subsubsection{Displacement statistics}

To study the dispersion behaviour of the displacement front, we quantify the displacement mean and variance. To this end, we define the $n$-th displacement moment in a streamwise direction as 

\begin{equation}
   m_n(t) = \langle [x(t) - x_0]^n \rangle, \label{eq:moment}
\end{equation}

where the angle brackets denote the average over all the particles and medium realizations. The displacement mean and variance are defined in terms of the first and second displacement moments as 

\begin{align}
   Z(t)  &= \overline{m_1}(t) \label{eq:com_def}\\
   \sigma^2(t) &= \overline{m_2}(t) - \left[\overline{m_1}(t)\right]^2. \label{eq:var_def}
\end{align}

The ensemble variance measures the spread of the mixing interface or the area swept by it. 
Furthermore, for purely advective transport, we consider the length of the displacement front, which is defined by 

\begin{align}
 \label{eq:ell_def}
   \ell(t) = \sum_i \sqrt{\Delta x_i(t)^2 + \Delta y_i(t)^2},
\end{align}

where $\Delta x_i(t) = x_i(t,\va + \Delta \va) - x_i(t,\va)$ and $\Delta y_i(t)$ accordingly. The sum runs over all line segments. For a homogeneous medium, $\ell(t) = \ell_0$ remains constant. Furthermore, we consider the dispersion of particle pairs initially separated by a small distance $|\Delta \va|$. Streamwise pair dispersion is quantified by 

\begin{align}
    \sigma_{\mathrm{eff}}^2(t) &= \frac{1}{2}\overline{\langle [x(t|\mathbf{a}) - x(t|\mathbf{a}+\Delta \mathbf{a})]^2 \rangle}.
    \label{eq:var_eff_def}
\end{align}

The effective variance is a measure of the width of the mixing
interface~\cite{Kitanidis1988, Dentz2023}.  

\subsubsection{Concentration statistics}

To quantify the volume occupied by the solute, we use the dilution index~\cite{Kitanidis1994}, which is defined by

\begin{align}
   E(t) = \overline{\exp[H(t)]}, && H(t) = -\int_{\Omega} d\vx \,c(\vx,t) \ln[c(\vx,t)]. 
   \label{eq:mixvol_def}
\end{align}

The probability density function (PDF) of concentration point values is defined by

\begin{equation}
   \Pi(c,t)=\frac{1}{V}\int_{\Omega} d\mathbf{r} \, \delta\left[c-c(\vx+\mathbf{r},t)\right],
   \label{eq:pdf_def}
\end{equation}

where $V$ is the volume of the domain $\Omega$. 

The concentration distribution for transport in an equivalent homogeneous medium
can be approximated by the Gaussian concentration distribution \cite{Pool2016}

\begin{align}
    c_0(\vx,t) = \frac{1}{\ell_0} \frac{\exp\left[-\frac{[x-z_0(t)]^2}{2 \sigma^2(t)}\right]}{\sqrt{2\pi\sigma^2(t)}},
    \label{eq:c_ana}
\end{align}

where the expressions for $z_0(t)$ and $\sigma_e^2(t)$ can be found in~\ref{app:hom}. The dilution index for this concentration distribution
is given by

\begin{align}
E_0(t) = \ell_0 \sqrt{2e \pi \sigma_e^2(t)},
\end{align}

where $e$ is the Euler constant. Based on this expression, we define a
``mixing`` variance $\sigma_{\rm mix}^2(t)$ in terms of the dilution index
$E(t)$ as

\begin{align}
\sigma_{\rm mix}^2(t) = \frac{1}{2 e \pi}\left[\frac{E(t)}{\ell_0}\right]^2. 
\end{align}

It describes the variance of a Gaussian distribution that occupies the same volume
as the actual concentration distribution. 

\section{Results and discussion}

\begin{figure}[htbp]
  \includegraphics[width=\linewidth, pagebox=artbox]{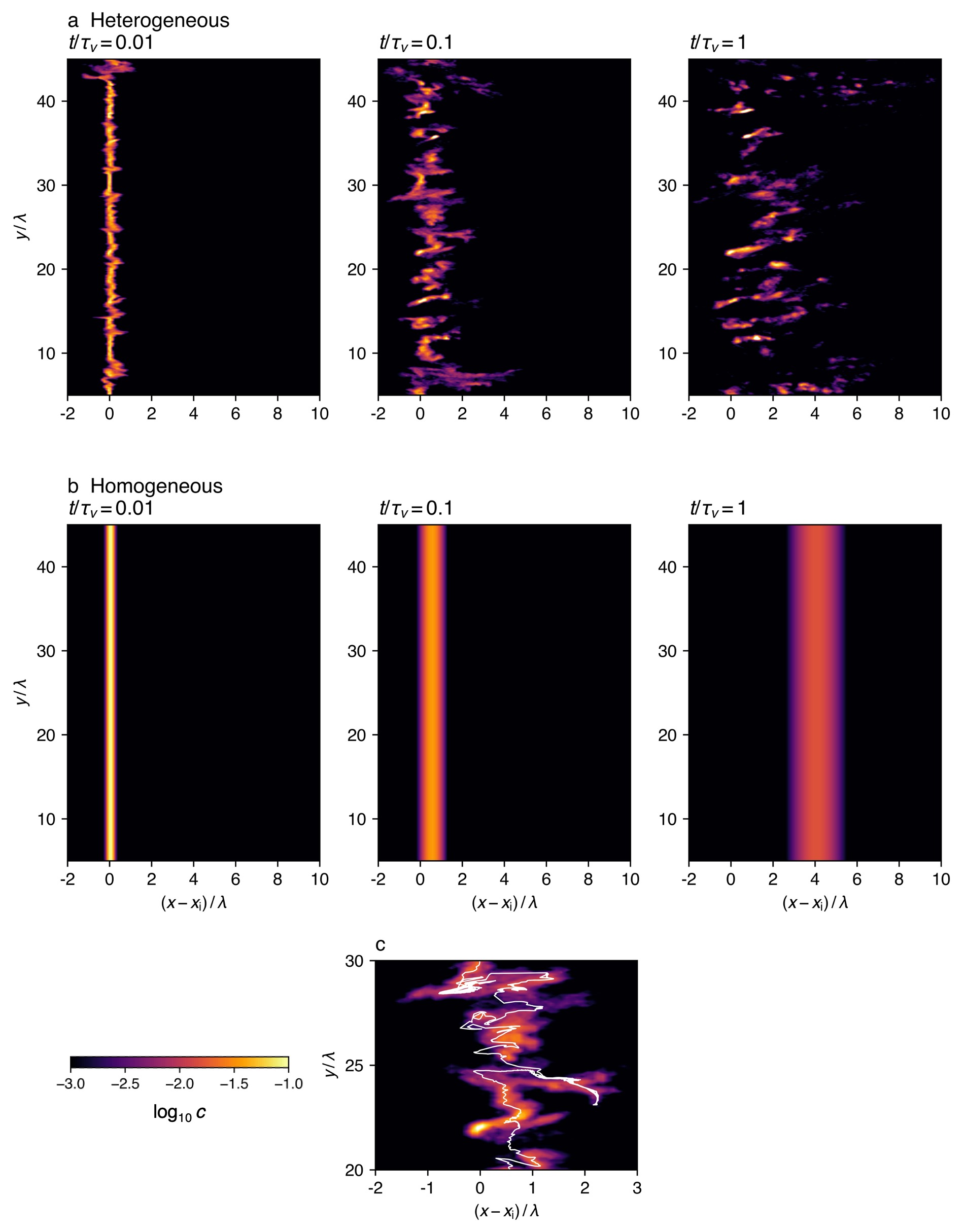}
  \caption{Temporal snapshots of concentration ($c$) distributions from (a)
    heterogeneous simulation results ($\sigma_f^2=2.0$) and (b) homogeneous
    analytical solutions (Equation\protect~\ref{eq:c_ana}) for
    $Pe=1000$. (c)
    The concentration distribution for the heterogeneous medium with $\sigma_f^2=2.0$ at $t/\tau=0.1$ is superposed by the purely
    advective line (shown by white).}
  \label{fig:conc_norm}
\end{figure}

We first consider purely advective transport to understand the stirring
properties of the underlying flow field. To this end, we discuss the time
evolution of the displacement mean and ensemble variance, and of the line
length. Furthermore, we analyze Poincar\'e maps of the flow field, which give
insight into the flow topology and its impact on the observed dispersion and
mixing patterns. Second, we study dispersive transport and mixing to
elucidate the interplay of temporal variability, medium compressibility and
spatial heterogeneity on solute mixing. Figure~\ref{fig:conc_norm} shows the
concentration distribution evolving from an initial lines source for homogeneous
and heterogeneous media.

\subsection{Purely advective transport and stirring}

\begin{figure}[htbp]
  \includegraphics[width=\linewidth, pagebox=artbox]{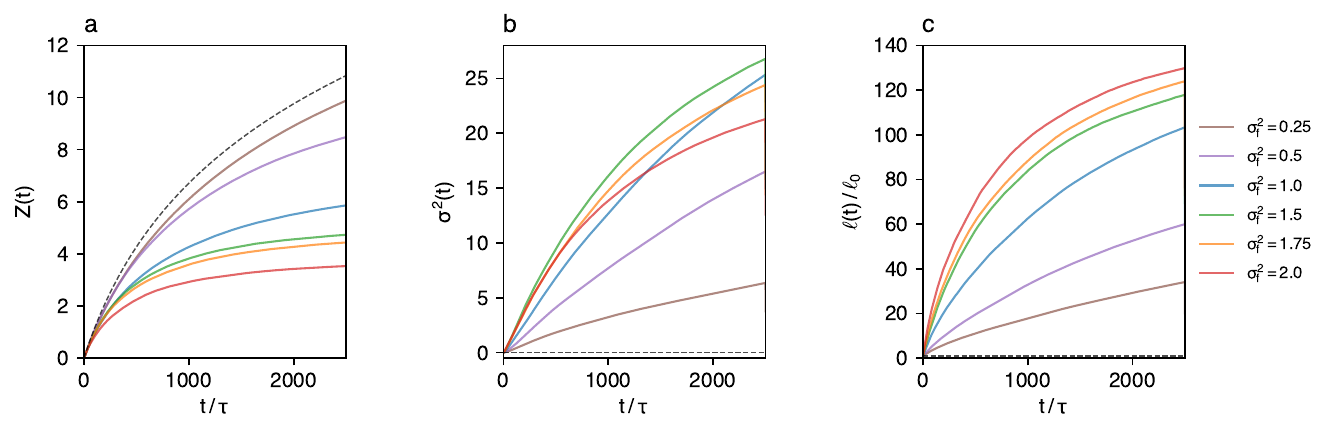}
  \caption{Temporal evolutions of period-averaged (a) centre of mass
    ($Z(t)$), (b) ensemble variance ($\sigma^2(t)$), and (c) displacement
    front length ($\ell(t)$) for various cases of $\ln K$ variance
    ($\sigma_f^2$).
    All cases
    are purely advective ($Pe=\infty$). Solid lines depict ensembled direct
    simulation results for heterogeneous media, and
    dashed lines are analytical expressions for homogeneous media
    ($\sigma_f^2=0$).}
  \label{fig:moment}
\end{figure}

Figure~\ref{fig:moment} shows the temporal evolution of the displacement mean, variance, and line length for different heterogeneity strengths $\sigma_f^2$. 
Figure~\ref{fig:moment}a shows that increasing
$\sigma_f^2$ leads to a delay in the advance of the centre of mass $Z(t)$
compared to a homogeneous medium.

\begin{figure}[htbp]
  \includegraphics[width=\linewidth, pagebox=artbox]{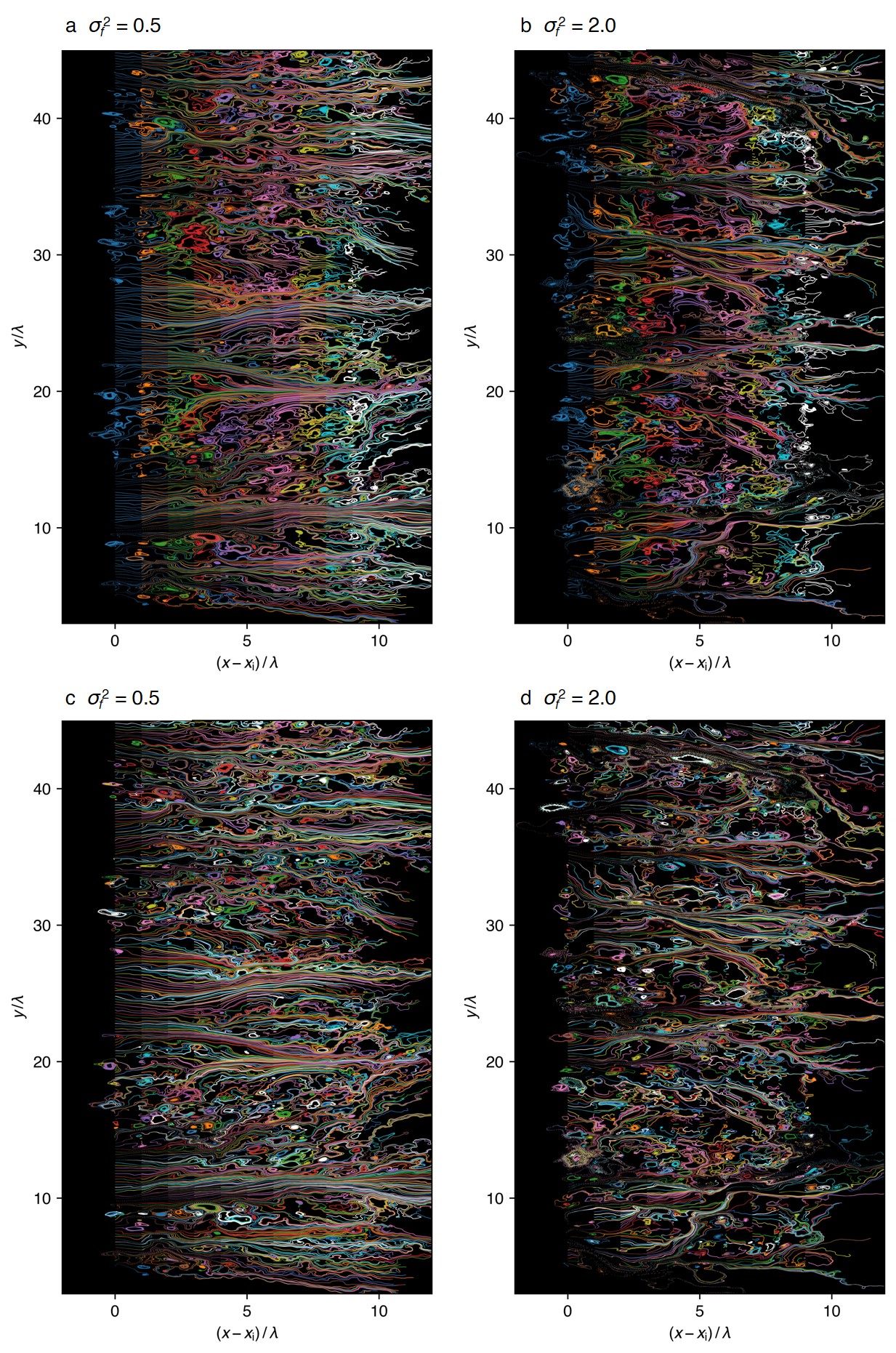}
  \caption{Poincaré sections for (a, c) $\sigma_f^2=0.5$ and (b, d)
    $\sigma_f^2=2.0$. All cases are purely advective ($Pe=\infty$). 
    Particles are initially seeded at
    interval of $\Delta x/\lambda = 1, \Delta y/\lambda = 0.2$, and coloured
    dots are particle potisions at $t/\tau = 1, 2, 3, \dots, 1500$. Panels with
    the same $\sigma_f^2$ show the same result with different colour schemes for
    initial particle positions along: (a, b) $x$- and (c, d) $y$-axis.}
  \label{fig:poincare}
\end{figure}

Figures~\ref{fig:moment}b and c show that $\sigma^2(t)$ and $\ell(t)$ increase
with time for heterogeneous media, while they are constant for equivalent
homogeneous media. This is due to the heterogeneity-induced distortion of the line
as shown in Figure \ref{fig:conc_norm}. With increasing $\sigma_f^2 < 1.5$, the
ensemble variance, which measures the area swept by the line, increases as expected from macrodispersion theory for uniform flow
\cite<e.g.>{Gelhar1983, Dagan1982}.

However, for $\sigma_f^2 \geq 1.5$, a discernible reversal of trends is observed
for times $\tau \gtrsim \tau_v$. Further increasing $\sigma_f^2$ leads to a
slowing down of the temporal evolution of $\sigma^2(t)$ compared to the lower
$\sigma_f^2$ cases. Thus, the effective area swept grows slower than for weak
heterogeneity. The line length $\ell(t)$ in contrast increases steadily with
increasing $\sigma_f^2$. The line is stretched and deformed into a lamellar
structure as shown in Figure~\ref{fig:conc_norm}. At early times $t < \tau_v$,
$\sigma(t)$ and $\ell(t)$ both are measures for the average interface length due
to flow deformation. For large times, however, $\sigma(t)$ quantifies the
extension of the area that is swept by the interface, that is, the support of
the distorted line. The interface length increases by stretching and folding
back on itself on a support whose growth rate decreases as illustrated in
Figure~\ref{fig:conc_norm}.
That is, the increase of the tortuosity of the line by stretching and folding
back on itself is stronger than the relative confinement due to the slowing down
of the centre of mass position and ensemble variance.

In summary, increasing heterogeneity in porous media decelerates the advancement of the displacement front, that is, it leads to relative confinement of the particle plume and enhances line stretching and folding. 
These observations have
implications for the understanding of mixing in coastal aquifers, but also in the
context of the design of groundwater remediation measures by reactant injection
and mixing, as well as geological storage.

To understand these behaviours, we consider Poincar\'e maps of the
underlying flow fields~\cite{lester2024}. Figure~\ref{fig:poincare} shows snapshots of the positions of advectively transported particles at times $t/\tau =
1-1500$ for two different heterogeneity variances. Particles are initially
seeded on a regular grid with spacing $0.2 \lambda$ over the flow domain. The upper panels use the same colours for particles initialized in the same column, and the lower panels in the same row, to illustrate the quenching of particles originating from different horizontal or vertical positions in chaotic flow regions. We observe confined stable regions of particles that move on
closed orbit, regular open trajectories, as well as chaotic regions. For
$\sigma_f^2 = 2$, we observe less regular open trajectories and an increase in
stable islands and chaotic regions compared to the case of weaker heterogeneity
with $\sigma_f^2 = 0.5$. For an equivalent homogeneous medium, all trajectories are of course straight lines. The retention of particles in stable and confined
chaotic regions explains the observations of Figure~\ref{fig:moment}. The
displacement mean slows down due to increased retention at increasing
$\sigma_f^2$. Similarly, the ensemble variance first increases with
heterogeneity compared to a homogeneous media but then slows down again due to particle retention. The interface length on the other hand increases due to
stretching and folding around the stable regions in subsequent flow periods.

\subsection{Dispersive transport and mixing}

\begin{figure}[htbp]
  \includegraphics[width=\linewidth, pagebox=artbox]{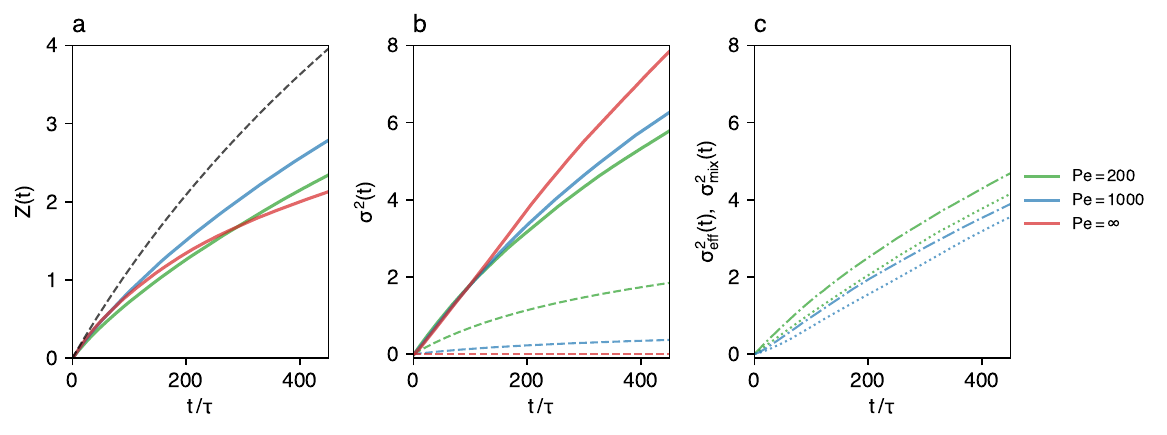}
  \caption{Temporal evolutions of period-averaged (a) centre of mass
    ($Z(t)$), (b) ensemble ($\sigma^2(t)$), and (c) effective
    (dash-dotted; $\sigma_{\mathrm{eff}}^2(t)$) and mixing (dotted;
    $\sigma_{\mathrm{mix}}^2(t)$) variances for various cases of
    P\'{e}clet number ($Pe$)
     Solid
    lines are from heterogeneous simulations ($\sigma_f^2=2.0$), and dashed
    lines in a and b are homogeneous analytical expressions.}
  \label{fig:var_comp}
\end{figure}

In this section, we study solute dispersion and mixing in the presence of local
scale dispersion, that is for finite local scale $Pe$. Figure~\ref{fig:var_comp}a
shows that the mean front displacement is larger for the high than for low $Pe$, while the mean displacement for the equivalent homogeneous medium is not affected by $Pe$. At large times, the displacements for finite $Pe$ are both larger than for the purely advective case. This behaviour can be explained by the fact that finite dispersion can release particles from stable and confined regions, and promote transport in regular regions. Thus, the displacement mean first increases for finite $Pe$. However, when $Pe$ is further increased, it decreases again because particles can be more efficiently been trapped in stagnant regions, which again leads to a retention of the plume. 

Figure~\ref{fig:var_comp}b shows that $\sigma^2(t)$ increases for increasing
$Pe$. Recall that the origin of dispersion is persistent velocity contrast
experienced by particles along pathlines. For small $Pe$, these contrasts
are rapidly smoothed out by transverse dispersion compared to large $Pe$,
similar in fact to Taylor dispersion in a tube. For equivalent homogeneous media, this is different because the variance scales with the local scale
dispersivity. In fact, for infinite $Pe$, $\sigma^2(t) = 0$.
Figure~\ref{fig:var_comp}c shows the evolution of the effective and mixing variances for different $Pe$. Both are smaller than the corresponding ensemble variance, and the mixing variance is the smallest measure. Yet, both are significantly larger than the variance for an equivalent homogeneous medium. 
Unlike the ensemble variance, they follow a similar trend as the variance for
the equivalent homogeneous media in that they decrease for increasing
$Pe$. Recall that the ensemble variance quantifies the area swept by the
solute line rather than the actual mixing area, and its magnitude is determined
by the spreading of the line. The effective variance quantifies the average
spreading of a point source, and the mixing variance is
an actual measure for the extension of the mixing volume, which is discussed
below. Thus, both measures are determined by local dispersion, especially at
small times where dispersion is the only mechanism to increase the mixing are. 

\begin{figure}[htbp]
  \includegraphics[width=0.7\linewidth, pagebox=artbox]{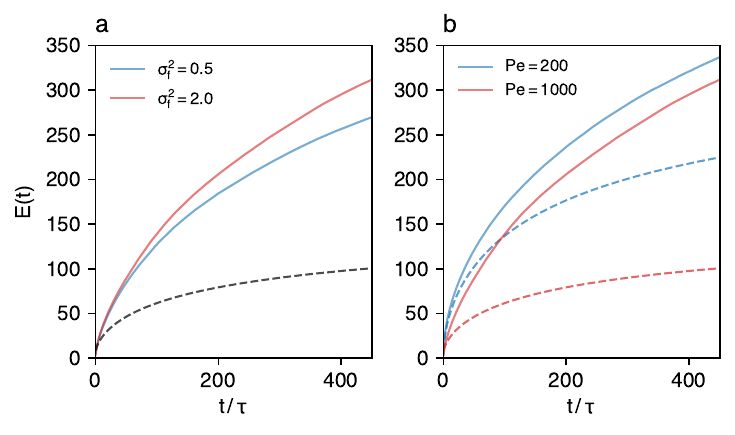}
  \caption{Temporal evolutions of period-averaged dilution index ($E(t)$) for
    various cases of (a) $\ln K$ variance ($\sigma_f^2$), (b) P\'{e}clet number
    ($Pe$).
    Unless indicated otherwise in the legend, the
    parameter values are $\sigma_f^2=2.0$ and $Pe=1.0\times10^3$
    Solid lines are from heterogeneous simulations, and
    dashed lines from homogeneous analytical expressions.}
  \label{fig:alpha_E}
\end{figure}

Figure~\ref{fig:alpha_E} shows the dilution index, that is, the mixing area, for
different heterogeneity variances and P\'eclet numbers. 
Figure~\ref{fig:alpha_E}a shows a dramatic increase in the mixing volume
in the heterogeneous media compared to the equivalent homogeneous media. This increase is attributed to the significant stirring action by the heterogeneous medium illustrated in Figure~\ref{fig:conc_norm}, which manifests the increase of the line length shown in Figure \ref{fig:moment}. As for the effective and mixing variance, we observe an increase in the dilution index with decreasing $Pe$. Thus, a decrease of the mean displacement, which indicates containment
of the plume, is concomitant with an increase of the mixing volume.

\begin{figure}[htbp]
  \includegraphics[width=\linewidth, pagebox=artbox]{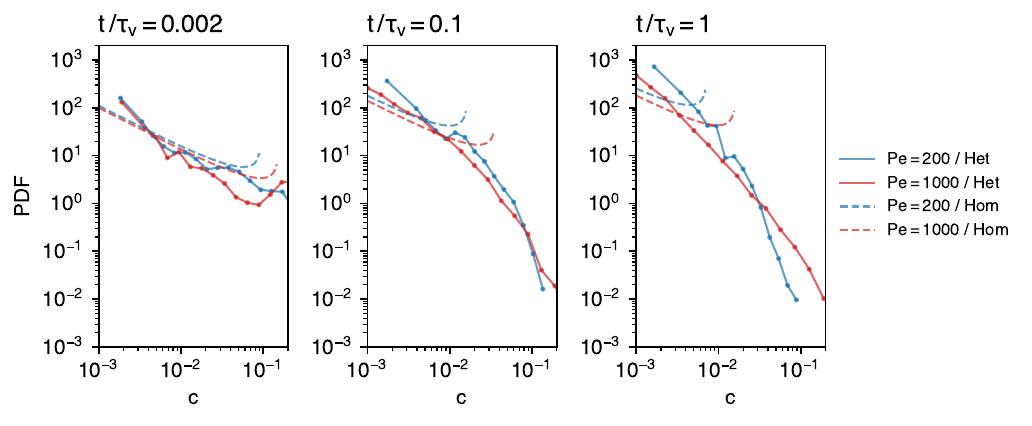}
  \caption{Temporal snapshots for probability density functions (PDFs) of concentration ($c$). For all cases $\kappa=2.1$ 
  $S_s=1\times10^{-2}$. Het: numerical results for heterogenous cases ($\sigma_f^2=2.0$); Hom: analytical solutions for homogeneous cases.}
  \label{fig:conc_norm_pdf}
\end{figure}

Figure \ref{fig:conc_norm_pdf} shows concentration point PDFs for $\sigma_f^2 =
1$ and two different $Pe$ compared to the behaviour expected for an equivalent
homogeneous medium. For homogeneous media, the PDF at small $Pe$ is generally
shifted toward smaller concentration values compared to large $Pe$. For the
heterogeneous case, we observe the same trend because local scale dispersion is
the ultimate mixing mechanism. Furthermore, at short times, the concentration PDF is expected to behave as in a homogeneous medium because mixing is primarily due to local scale dispersion, which indeed varies along the line source according to the spatially variable flow velocity. Recall that dispersion is proportional to the
velocity magnitude. Thus, at short times, the PDFs for the heterogeneous and homogeneous media are qualitatively similar. At late times, however, the concentration PDFs for the heterogeneous media show a behaviour markedly different from homogeneous media. The concentration support is larger than for
homogeneous media, specifically toward larger concentrations. This can be
attributed to weak mixing in low-velocity zones, for which local scale
dispersion is small. The probability of low concentration values on the other
hand is increased compared to the homogeneous media as a consequence of increased
stirring and mixing action that is also reflected in the behaviors of the
effective and mixing variances and the dilution index. The concentration PDFs for low and high P\'eclet numbers intersect with increasing time because for low $Pe$ the probability of high concentration values decreases faster than for high $Pe$ and at the same time the probability for low values increases faster.

\section{Conclusions}

In conclusion, our results show that the interaction of medium heterogeneity, compressibility and temporal flow fluctuations leads to a significant increase in mixing compared to equivalent homogeneous media. While
increased heterogeneity leads to relative confinement of the solute plume due
to a slowing down of the growth rate of the mean displacement, it increases
spreading, stirring and ultimately solute mixing. Poincar\'e maps of the
fluctuating Lagrangian flow elucidate the origin of these behaviours, which lie in the creation of stable, regular and chaotic regions. In fact, for increasing heterogeneity, the relative proportion of these structures increases, which
leads to a decrease in the advance of the mixing front and the area swept by
it.

These findings have implications for a series of environmental and engineering
applications that involve periodically forced flow in porous media. This
includes the evaluation of the mixing of salt and freshwater in coastal aquifers
under tidal forcing, as well as the assessment of groundwater remediation
measures due to the periodic injection of reactants into contaminated soils or
groundwater. Furthermore, our results can guide geostorage efforts that rely on
the minimization of mixing with the ambient fluid, which is key in underground
hydrogen storage, or its maximization, which is desired in geological carbon
dioxide storage.

\appendix
\section{Numerical implementation of particle tracking}
\label{app:numerical}
By applying the It\^o–Taylor integration scheme to the advection-dispersion equation, we write the discrete-time displacement of a particle for the case of constant porosity and isotropic dispersion as~\cite{Labolle2000}

\begin{eqnarray}
    \vx(t+\Delta t) &=& \vx(t) + \vu\left(\vx, t+\Delta t\right)\Delta t + \mathbf{B}\left(\vx+\Delta \vx, t+\Delta t\right)\boldsymbol{\zeta}(t)\sqrt{\Delta t} \label{eq:RWPT}\\
    \Delta \vx &=& \mathbf{B}\left(\vx, t+\Delta t\right)\boldsymbol{\zeta}(t)\sqrt{\Delta t}
\end{eqnarray}

where $\Delta t$ is the time step, $\vu(\vx, t) = \vq(\vx, t)/\phi$ is the Lagrangian velocity, and $\boldsymbol{\zeta}(t)$ is a Gaussian white noise with zero mean and a correlation function $\langle\zeta_i(t)\zeta_j(t')\rangle = \delta_{ij} \delta(t - t')$. For simplicity, we assume an isotropic $\mathbf{D}$ with zero molecular diffusion ($D_{\mathrm m}=0$): $D_{ij} = \alpha |\vu|\delta_{ij}$. Under this assumption, the displacement matrix $\mathbf{B}$ is written as~\cite{Lichtner2002, Salamon2006}

\begin{equation}
    \mathbf{B}=
    \begin{bmatrix}
    u_x\sqrt{2\alpha/|\vu|} & -u_y\sqrt{2\alpha/|\vu|}\\
    u_y\sqrt{2\alpha/|\vu|} & u_x\sqrt{2\alpha/|\vu|}
    \end{bmatrix}.
\end{equation}

Transport is simulated by particle tracking with our Fortran code (see Text S2 for code validation). 

For calculating the particle distribution metrics for purely advective cases, $N_{\mathrm s} = 4,000$ particles are seeded at $t=T$ along the line $x = 5\lambda (=x_{\mathrm i}), 5\lambda \leq y \leq 45\lambda$ at a constant interval of $\Delta y = 1\times10^{-2}\lambda$. This number of particles is sufficiently large to ensure convergence~\cite{deDreuzy2007}. See Appendix A in~\citeA{TajimaTED} for details on velocity field interpolation and an adaptive time step control. We ran $N_{\mathrm r} = 25$ simulations with different realizations of heterogeneous $K$ fields for each case.

To calculate concentration and the related observables for cases with finite local diffusion, we performed an additional simulation ($N_{\mathrm r} = 1$) for each case with an increased number of particles, $N_{\mathrm s} = 100,000$, with $\Delta y = 4\times10^{-4}\lambda$. The concentration was calculated from the obtained particle distributions using a Gaussian kernel density estimator~\cite{fernandez2011optimal}.

\section{Effective transport in a homogeneous porous medium}
\label{app:hom}
For the convenience of the reader, we summarize here the results for effective
transport in a homogeneous porous medium from \citeA{Pool2016}. These authors
used a two-scale expansion to quantify the transport behaviour of the
time-averaged concentration distribution. The time-averaged concentration
$\overline c_0(x,t)$ satisfies the advection-dispersion equation

\begin{align}
\frac{\partial \overline{c}_0(x,t)}{\partial t} + \frac{\partial}{\partial x} v_e(x) \overline{c}_0(x,t) - \frac{\partial^2}{\partial x^2} [D_e(x) + D_0] \overline{c}_0(x,t) = 0,
\end{align}

which describes solute transport for $t \gg \tau$. The effective transport velocity and dispersion coefficient are given by

\begin{align}
v_e(x) = \frac{v_0^2 \mu \tau \exp(- \mu x)}{4 \pi}, && D_e(x) = \frac{2 \alpha v_0}{\pi} \exp(- \mu x). 
\end{align}

Based on this effective formulation, \citeA{Pool2016} derive the following approximate solutions for the centre of mass position and displacement variance, 

\begin{align}
  Z(t) &= \frac{1}{2\mu}\ln\left(\frac{t}{\tau_v}+1\right)
  \label{eq:com}\\
  \sigma_e^2(t) &= \frac{\sigma^2_0}{(1 + t/\tau_v)^2} + \frac{4 D_{\mathrm e} \tau_v}{5}\left[\frac{(1 + t/\tau_v)^{\frac{5}{2}} - 1}{(1 + t/\tau_v)^2}\right] + 2 D_0 t \left[\frac{1}{3} + \frac{(t/\tau_v)^2 - 1}{3 t/\tau_v (1 + t/\tau_v)^2}\right],
  \label{eq:var}
\end{align}

where $\sigma_0^2 = \sigma^2(t = 0)$ is the initial variance. In this study, we
set $\sigma^2_0=0$ and $D_0=0$. These authors find that the concentration
distribution can be approximated by the one-dimensional Gaussian distribution

\begin{align}
    c(\vx,t) = \frac{1}{\ell_0}\frac{\exp\left(-\frac{[x-z_0(t)]^2}{2 \sigma^2(t)}\right)}{\sqrt{2\pi\sigma^2(t)}},
    \label{eq:c_ana}
\end{align}

where $\sigma_e^2(t)$ is given by Equation~\eqref{eq:var}.

For the Gaussian concentration distribution (Equation~\ref{eq:c_ana}), the
dilution index can be determined explicitly and is given by 

\begin{equation}
   E(t) = \ell_0\sqrt{2e\pi\sigma^2(t)}.
   \label{eq:E_ana}
\end{equation}

Similarly, one obtains for the concentration PDF the explicit analytical expression

\begin{equation}
   \Pi(c,t)=\frac{\mathbb I\left[c_{\mathrm{th}} < c \leq c_{\mathrm{max}}(t)\right]}{2c \sqrt{\ln\left[c_{\mathrm{max}}(t)/p\right]\ln\left[c_{\mathrm{max}}(t)/c_{\mathrm{th}}\right]}},
\end{equation}

where $\mathbb I(\cdot)$ is the indicator function, which is one if its argument is true and zero otherwise, $c_{\mathrm{th}}$ is the threshold concentration value, and $c_{\mathrm{max}}(t)$ is the maximum concentration in the domain at each time, given from Equation~\eqref{eq:c_ana} by

\begin{equation}
   c_{\mathrm{max}}(t)=\frac{1}{\ell_0\sqrt{2\pi\sigma^2(t)}}.
\end{equation}

\section{Concentration PDFs with pure diffusion}
\label{app:pure_diff}
Figure~\ref{fig:conc_norm_pdf} illustrates that the largest concentration value in the heterogeneous cases becomes higher than that in the homogeneous cases. We hypothesize that this result is attributed to the particle entrapment by low-velocity regions in heterogeneous media. To confirm this hypothesis, we consider additional cases with finite diffusion, which is constant and independent of velocity, instead of the velocity-dependent local dispersion assumed throughout this study. The dispersion tensor $\mathbf{D}$ here becomes

\begin{align}
    D_{ij} = D_0 \delta_{ij}
\end{align}

where $D_0 = \alpha v_0$.

\begin{figure}[htbp]
  \includegraphics[width=\linewidth, pagebox=artbox]{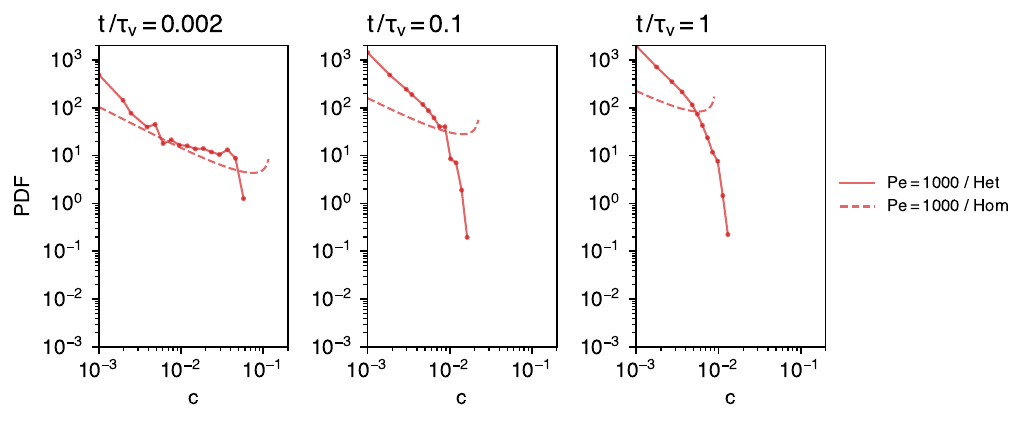}
  \caption{Temporal snapshots for probability density functions (PDFs) of concentration ($c$) with pure diffusion. For all cases $Pe=1000$. Het: numerical results for heterogenous cases ($\sigma_f^2=2.0$); Hom: analytical solutions for homogeneous cases.}
  \label{fig:conc_purediff_pdf}
\end{figure}

Figure~\ref{fig:conc_purediff_pdf} shows that the maximum concentration value in the heterogeneous case is smaller than the homogeneous case, which is opposite from the cases with velocity-dependent local dispersion (Figure 6 in the main article). This observation suggests that particles can escape from the low-velocity regions in heterogeneous media with diffusion independent of velocity, corroborating the hypothesis above. Meanwhile, as observed in the dispersive cases, Figure~\ref{fig:conc_purediff_pdf} illustrates that dilution is enhanced in the heterogeneous case compared to the homogeneous case.

\acknowledgments
S. Tajima acknowledges the financial support by the JSPS KAKENHI Grant Number JP23KJ0431 (Grant-in-Aid for JSPS Fellows). The Authors thank Daniel Lester (RMIT University) for insightful discussions.

\bibliography{tide}
\end{document}